\documentclass{elsart}

\usepackage{graphicx}

\begin{document}
\begin{frontmatter}
\title {First study of radiation hardness of lead tungstate crystals
 at low temperatures}
\author[IHEP]{P.A.~Semenov},
\author[IHEP]{A.V.~Uzunian\thanksref{addr}},
\thanks[addr]{corresponding author, email: uzunian@sirius.ihep.su}
\author[IHEP]{A.M.~Davidenko},
\author[IHEP]{A.A.~Derevschikov},
\author[IHEP]{Y.M.~Goncharenko},
\author[IHEP]{V.A.~Kachanov},
\author[IHEP]{V.Y.~Khodyrev},
\author[IHEP]{A.P.~Meschanin},
\author[IHEP]{N.G.~Minaev},
\author[IHEP]{V.V.~Mochalov},
\author[IHEP]{Y.M.~Melnick},
\author[IHEP]{A.V.~Ryazantsev},
\author[IHEP]{A.N.~Vasiliev},
\author[RRC]{S.F.~Burachas},
\author[RRC]{M.S.~Ippolitov},
\author[RRC]{V.~Manko},
\author[RRC]{A.A.~Vasiliev},
\author[MEPhI]{A.V.~Mochalov},
\author[GERM]{R.~Novotny},
\author[VILNUS]{G.~Tamulaitis}

\date{\today}

\address[IHEP]{Institute for High Energy Physics, Protvino, Russia}
\address[RRC]{RRC,''Kurchatov Institute'',Moscow, Russia}
\address[MEPhI]{Moscow Engineering Physics Institute (State University),
 Moscow, Russia}
\address[GERM] {Justus Liebig-Universitaet, Giessen,
 II Physikalisches Institut, Germany}
\address[VILNUS]{Vilnius University, Vilnius, Lithuania}

\begin{abstract}
The electromagnetic calorimeter of PANDA at the FAIR facility
will rely on an operation of lead tungstate (PWO) scintillation crystals
at temperatures near  -25~$^\circ$C to provide sufficient resolution
for photons in the energy range from 8~GeV down to 10~MeV.
Radiation hardness of PWO  crystals was studied at the IHEP (Protvino) 
irradiation facility in the temperature range from room temperature down
 to -25~$^\circ$C.
These studies have indicated a significantly different behaviour
in the time evolution of the damaging processes well below
room temperature. 
Different signal loss levels at the same dose rate, but
at different temperatures were observed. The effect of a deep
suppression of the crystal recovery process at temperatures below
  0~$^\circ$C has been seen.

\end{abstract}
\end{frontmatter}

\section{Introduction}

Total absorption shower counters made of lead tungstate ($PbWO_4$ or PWO)
 scintillation crystals are very promising for development of high quality
 electromagnetic calorimeters\cite{first}.
 They have already been used in the PRIMEX
 experiment at JLaB and manifested a brilliant $\pi^0$-mass resolution
 ($1.5 MeV/c^2$) \cite{Mochalov}.
 The CMS electromagnetic calorimeter
 consisting of 62,000 PWO crystals in the barrel part and 15,000 
 in the endcap part is due to start operation at LHC soon \cite{cms}.
 Both the calorimeters work at room temperature.
ALICE, another calorimeter at LHC, will be the first PWO-based
 calorimeter in the world operating at a temperature decreased
 down to -25~$^\circ$C.
An increase of the light yield by a factor of 3 in comparison with
 operation at room temperature (+20~$^\circ$C) is expected~\cite{alice}.
Moreover, the noise of
 the calorimeter electronics will be reduced. Both these effects will 
  improve the energy resolution of the calorimeter.
 The same operation temperature of -25~$^\circ$C  is selected
 also for 20,000 PWO crystals in the PANDA  calorimeter, a new
 experiment in the FAIR project at GSI, Germany~\cite{panda} . 
Radiation hardness is another important issue in application
 of PWO scintillators in high energy physics experiments.
It is well known that crystal scintillators suffer from radiation damage.
 This problem has been extensively studied over the last seven years
 at CMS~\cite{ann,zhu} and BTeV, the Fermilab experiment
 which was terminated in 2005~\cite{nim2,nim3,nim7}.
 However, all the studies of PWO radiation hardness have been carried out
 only at room temperature.
 
In this paper we present the results of the first investigation
 of PWO crystal radiation hardness at temperatures below 0~$^\circ$C.
 The temperature
 was varied in the range from +20~$^\circ$C  down to -25~$^\circ$C.
 The study was carried out for two different radiation loads of 20 and
2~rad/h (radiation loads for the PANDA endcap electromagnetic calorimeter 
are estimated to be up to 2-3~rad/h).
 
 \begin{table}[b]
    \begin{center}
    \caption {PWO crystal specifications data before irradiation}
    \label{tab:summary}
      \begin{tabular}{|c||c|c|c|c|c|c|} \hline
Serial &Doped& Concen- &growth  & Light Yield & Trans-      & Trans-  \\
number &element&tration,&number & ($phe/MeV$)& mittance    & mittance \\
       &       & ($ppm$)&       &    & at 420 nm,(\%) & at 632 nm,(\%) \\
 \hline \hline
a15608 &Gd&80&1  & 10.9    &  69.0  & 73.9   \\ \hline
a15517 &Gd&80&1  & 10.5    & 71.3  &  74.0  \\ \hline
a15851 &Gd&80&1  & 11.0    & 71.3  &  73.9  \\ \hline \hline
a1412   &Y&100&5  & 9,2     &  64.3  &  72.6    \\ \hline
a1434  &Y&100&4   & 9.0    & 65.6  &  73.1   \\ \hline
a15500 &Y&90 &1    & 12.0    & 70.2   & 74.0  \\ \hline \hline
a16618 &La& 38&1  & 11.2    &  71.9  & 74.0   \\ \hline
a16628 &La/Y&22/20&1  & 12.3    &  71.3  & 74.0   \\ \hline \hline
b389   &N/A$^*$&N/A&N/A & 10.3    &  70.5  &  72.9    \\ \hline
b404   &N/A&N/A &N/A & 8.9    &  66.8  &  71.9  \\ \hline

      \end{tabular}
    \end{center}
{\small $*${Data not available}}
\end{table}

We used eight crystals produced by  the ''North Crystals''
 Company in Apatity 
 (denoted by the letter ``a'' in this paper)
 and two more crystals produced by the Bogoroditsk Techno-Chemical Plant
(``b''). 
The crystals  which are marked as ``b'', as well as a1412 and a1434
were produced at the end of 2002. The other crystals were produced in 2005.
 The dimensions of the crystals were $\sim 22\times 22$~mm$^2$
 in cross section and $180$ mm in length.
The crystal specifications data given by manufactures
 before the irradiation are provided in Table~\ref{tab:summary}.
The data show that these samples are typical examples of
 PWO scintillation crystals with similar transmittance
 and light yield however they differ by the doped elements
 and the growth number (number of a crystal growing in the same crucible
without cleaning of it).

\section{Experimental procedure }

\begin{figure}[h]
\centering
\includegraphics[width=80mm,height=75mm]{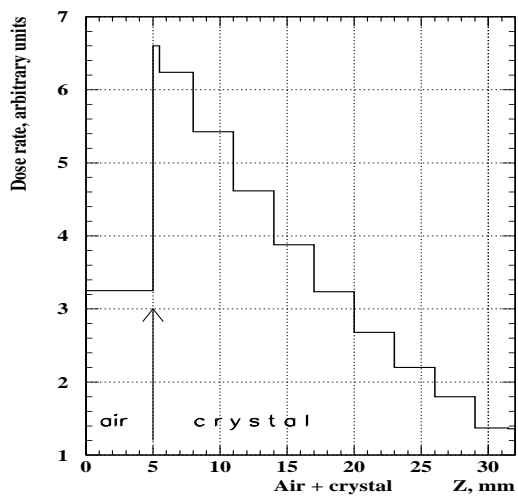}
\caption{ Transverse dose rate profile for crystal under
$^{137}Cs$ gamma irradiation.
The dose rate was estimated both in air and crystal, as indicated.}
\label{fig:cs137prof}
\end{figure}

A radioactive source $^{137}Cs$ emitting 661 KeV photons and having
activity of  $5\cdot 10^{12}$ Bq was used to irradiate the crystals.
Five crystals were irradiated simultaneously during each exposition.
 A commercial dosimeter, DKS-AT1123, was used to measure
 the dose rate (in air) in proximity to the crystals.
The accuracy is better than 20\%~\cite{http}.
  The simulated transverse radiation dose rate profile in crystal is shown 
 in Fig.~\ref{fig:cs137prof}~\cite{nim5}.

\begin{figure}
\centering
\includegraphics[width=110mm]{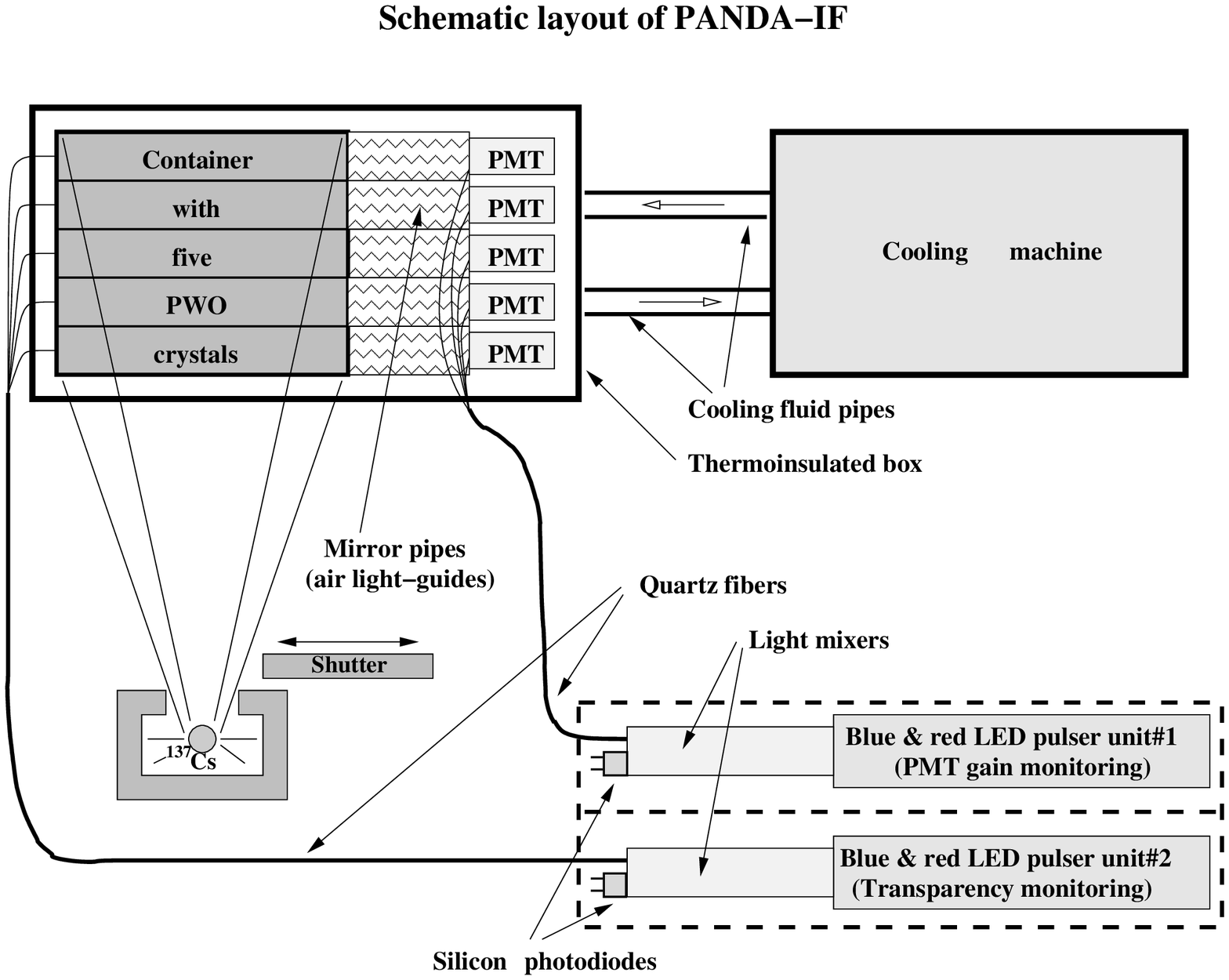}
\caption{Layout of the gamma irradiation setup. The details are
 in the text.}
\label{fig:pandaif}
\end{figure}

Layout of the gamma irradiation setup is depicted in 
Fig.~\ref{fig:pandaif}. 
Cryothermostat LAUDA RC6CP was exploited to stabilize and control
 the temperature of the crystal set under irradiation. 
  Cooling power of $300 W$ provided by this
cryothermostat was sufficient to maintain the constant temperature
for a long time with the absolute accuracy of  0.1~$^\circ$C.
 Glycol aqueous solution circulating between the LAUDA bath 
and the heat exchanger plate, where the crystals were installed,
 was used to maintain the crystals at a required temperature.
The bases of the photomultiplier tubes (PMTs), which were installed outside
 the heat exchanger, were the main heat sources inside the crystal compartment.
To make the temperature distribution more uniform, a permanent gas
flow inside the compartment was provided using a fan. Dry nitrogen gas
 was supplied to the compartment to avoid problems with moisture gathering.
To monitor the temperature of the crystals, several thermosensors
 (type PT100 and PT1000) were installed on the heat exchanger plate
 and directly on the crystals.
To monitor variation of the light transmittance of the crystals
 due to irradiation as well as possible PMT gain instabilities,
 we used an LED-based monitoring system.
This system enabled us to investigate the radiation-induced absorption
 of the crystals in two different spectral regions, in the vicinity
 of 450 $nm$ and 640 $nm$.
 The system consisted of two similar modules.
 Each module contained blue or red LEDs with individual
 drivers, light mixer, one photodiode and fiber bundle.
 The system was similar to that described in more detail in~\cite{nim8}.

A direct current (DC) method was used to test the radiation hardness
 of the crystals. The mean current value of the PMT through a 10 kOhm
 resistor was measured by sigma-delta ADC. This signal was proportional
 to the luminescence intensity and, consequently, to the light output
 of the crystal. The change of this signal provided an information on
 the changes in the light output invoked by irradiation.
 Simultaneously, the LED based monitoring system provided information
 on variation in crystal light transmittance.
 Since the level of the DC signal was low
 (a few microamps), sensitive measurement
 devices were installed in a close proximity to the photodetectors.
 Digitized signals were transferred further to a remote computer
 using the RS485 protocol.

Ten crystals (two groups, five crystals in each group)  were irradiated
 at the dose rate of 20 rad/h (and 2 rad/h) at different temperatures.
 Each cycle lasted from 50 to 300 hours.
 Before each cycle, the crystals were exposed
 to the daylight at room temperature  to recover
 after the previous cycle of irradiation.
 The recovery was tested by a reference measurement of the light output
 at $+20^\circ C$. We observed that the crystals restore their
 initial light output within 5\% after four days of such recovery.  

\section{Results}

\begin{figure}[t]
\parbox[l]{0.48\hsize}{
\includegraphics[width=1.\hsize,height=60mm] {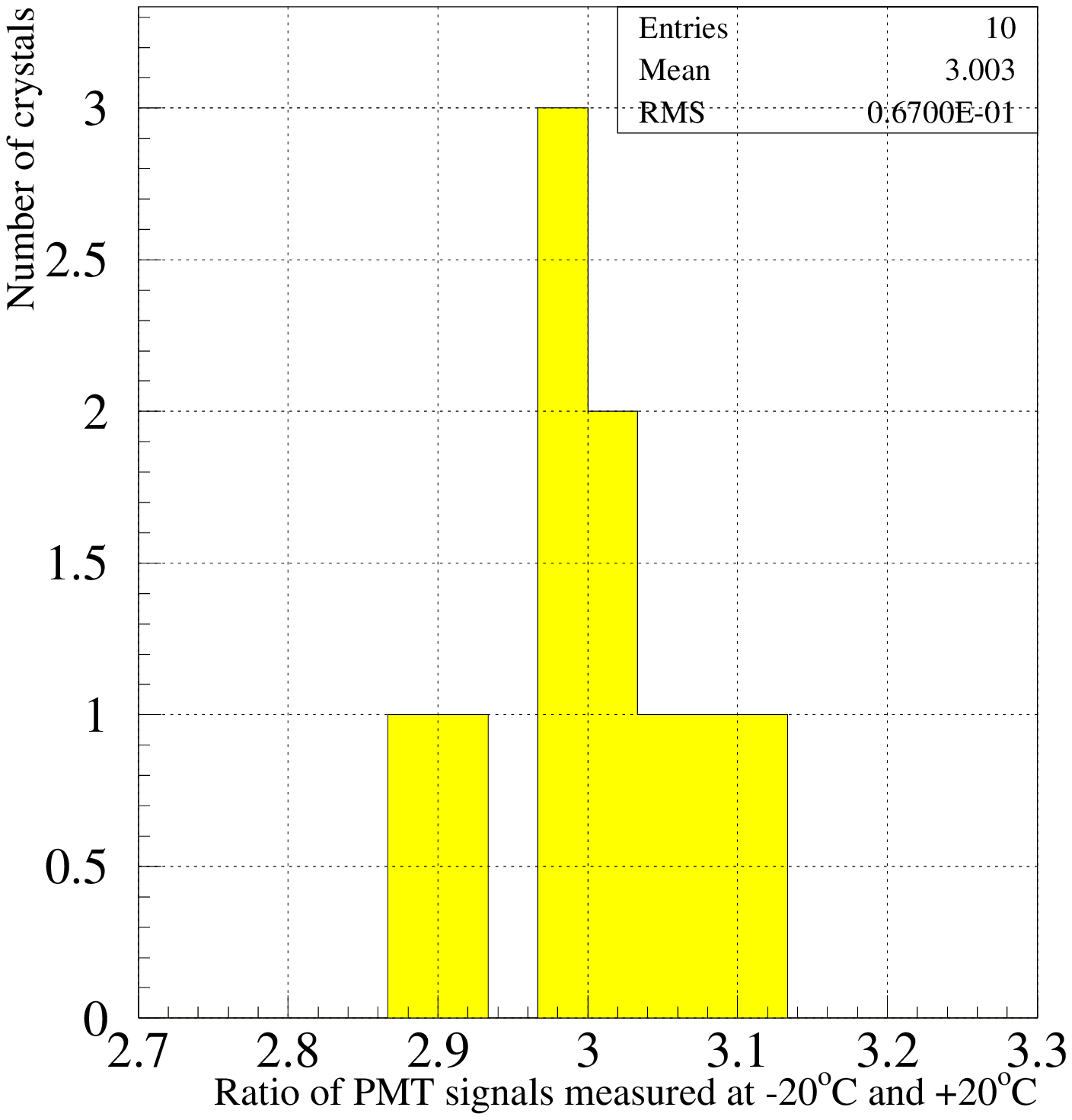}
\caption{Distribution of the ratio of PMT signals measured 
 at $-20^\circ C$ and $+20^\circ C$ among all ten crystals.}
\label{fig:jump_dist}
}
\hfill
\parbox[l]{0.48\hsize}{
\includegraphics[width=1.\hsize,height=55mm] {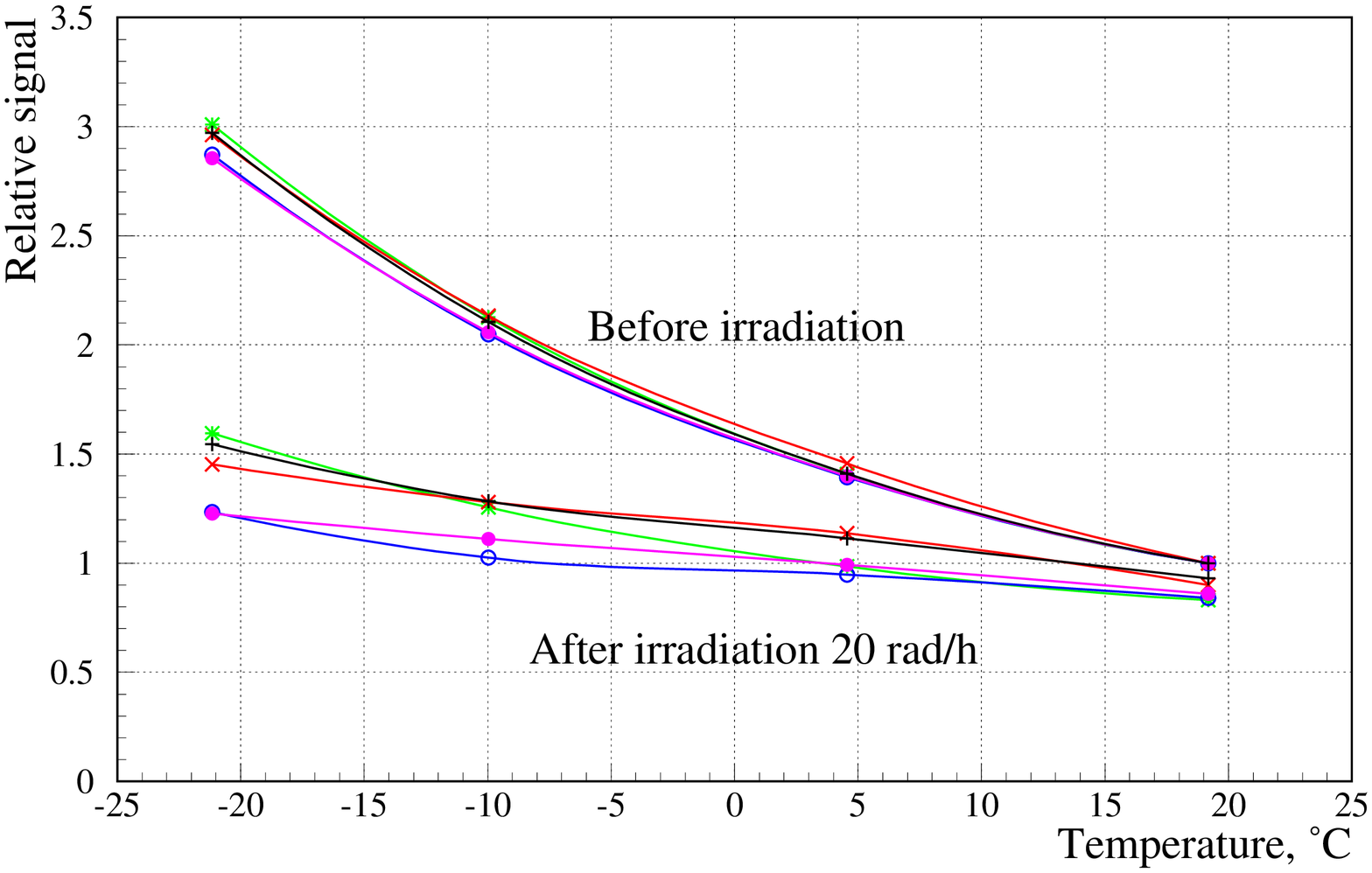}
\vspace*{-0.4cm}
\caption{Relative light output as a function of crystal temperature.
 Five PWO samples were measured before irradiation and after irradiation
 over 80 hours at the dose rate of 20 rad/h.}
\label{fig:temp_dep1}
}
\end{figure}

\begin{figure}[b]
\centering 
\includegraphics[width=80mm,height=70mm] {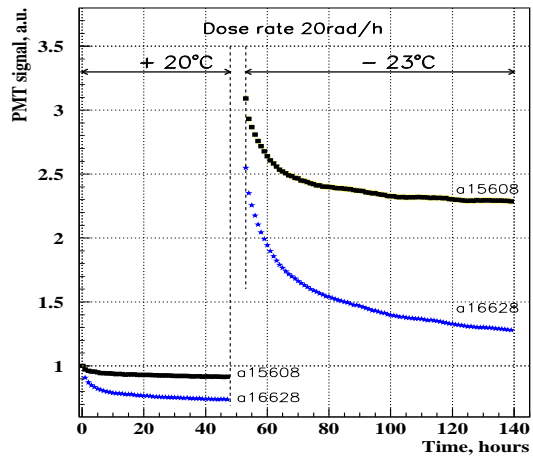}
\caption{Time evolution of PMT signal of two samples (a15608 and a16628)
 subjected to irradiation at dose rate of 20 rad/h.
 Periods when constant temperatures were maintained (+20~$^\circ$C and
  -23~$^\circ$C) are indicated by horizontal arrows.}
\label{fig:type20-23}
\end{figure}

As it might be expected, cooling of the scintillation crystals
increases their light output. Decrease of temperature
 from +20~$^\circ$C down to -20~$^\circ$C resulted in an increase
 in the PMT signal
 corresponding to the light output by approximately a factor of 3.
 The distribution of this enhancement factor for all 10 samples
 under study is presented in Fig.~\ref{fig:jump_dist}. 
The PMT current as a function of temperature for five crystals from
 one group is presented in Fig.~\ref{fig:temp_dep1}.
 Four cycles of measurements at four different temperatures,
 +20~$^\circ$C, +5~$^\circ$C, -10~$^\circ$C, and -22~$^\circ$C,
 were performed. The curves in Fig.~\ref{fig:temp_dep1} are just
 a guide for the eye.
 Before the irradiation, the light output at different temperatures
 of all five samples was the same within the experimental error.
 After the irradiation, the temperature dependence of the light output
 became less pronounced. The trend of this dependence was the same
 for all crystals, but the differences in absolute values 
were much higher than those before irradiation.
 As the radiation hardness of the crystals is slightly different,
 they loose different portions of their initial light output
 after the same period of irradiation at the same dose rate.

\begin{figure}[t]
\centering
\includegraphics[width=110mm]{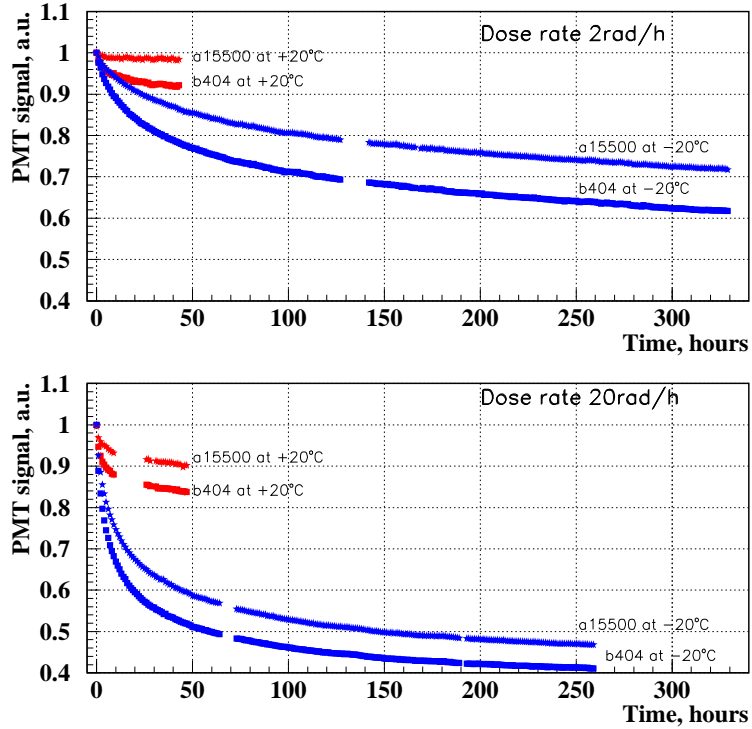}
\caption{Time evolution of external light yield of two PWO crystals
 measured under 2 and 20~rad/h irradiation dose rate at temperatures
 of -20~$^\circ$C and +20~$^\circ$C, as indicated.}
\label{fig:longterm+20-20}
\end{figure}

 To test the time evolution of the variation of the light output,
 we measured the PMT DC current as a function of time when
 the crystals were subjected to persistent irradiation at
 a constant dose rate. Typical curves of light output change under
 irradiation at the dose rate of 30 rad/h are shown
 in Fig.~\ref{fig:type20-23}. Initially, +20~$^\circ$C temperature
 was maintained. The signal dropped within the first 20 hours and
 approached practically a constant level afterwards.
 The drop was different for different crystals. As discussed above, decrease
 of temperature down to -23~$^\circ$C caused an increase of the signal by a
factor of approximately 3. The further irradiation caused a decrease
 of the signal. The signal also tends to a constant value.
 However, the time necessary to reach the constant value is longer,
 more than 300 hours, instead of 30 h at +20~$^\circ$C.
The next set of measurements was performed at the dose rate of 2 rad/h,
 which is about the maximal dose rate expected for the crystals
 in calorimeter for PANDA experiment.
 The same groups of the crystals were irradiated at +20~$^\circ$C 
 and -20~$^\circ$C. Fig.~\ref{fig:longterm+20-20} shows that
 after initial drop the signal in all expositions tends to reach a
 constant value. The higher is the dose rate, the lower is the constant
 level and the faster this constant level is reached.
 As demonstrated also in Fig.~\ref{fig:longterm+20-20}, decrease
 of temperature enhances the initial signal drop and delays
 the process of approaching a constant light output level.

\begin{figure}[p]
\centering
\includegraphics[width=100mm]{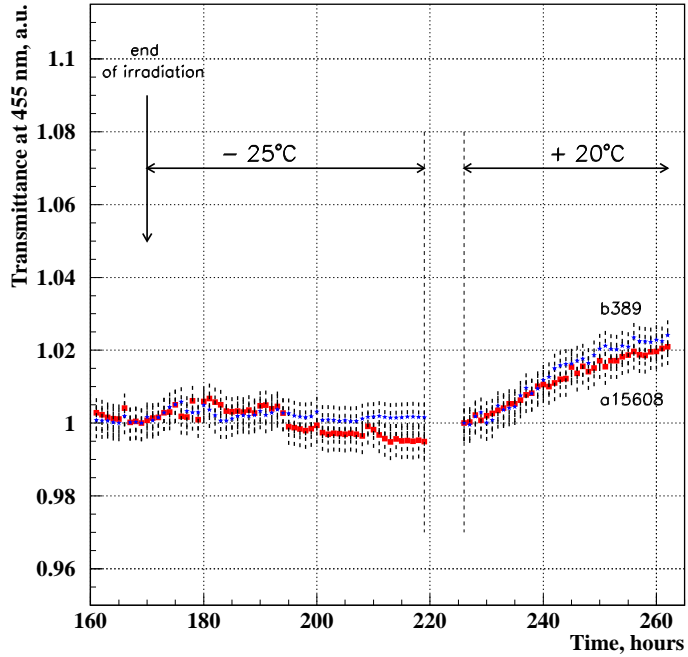}
\caption{Time evolution of transmittance at 450 nm after
 exposition to irradiation. End of irradiation is marked by
 a vertical arrow; the periods of constant temperatures
  (-25~$^\circ$C and +20~$^\circ$C) are marked by horizontal arrows.
The transmittance values at the beginning of recovery at -25~$^\circ$C
 and +20~$^\circ$C were normalized to one.}
\label{fig:recovery}
\end{figure}

 \begin{table}[p]
    \begin{center}
    \caption {Decrease of PMT DC signal in [\%] as a result of irradiation
 over 80 hours at the two dose rates and the two temperatures
 for the ten crystals}
    \label{tab:summary2}
      \begin{tabular}{|c||c|c||c|c|} \hline
Serial &2 rad/h         & 2 rad/h &  20 rad/h         & 20 rad/h  \\
number &at +20~$^\circ$C & at -20~$^\circ$C & at +20~$^\circ$C & at -20~$^\circ$C \\  
 \hline \hline
a15608 &2 &13      &9  & 30    \\ \hline
a15517 &1 &16      &8  & 43    \\ \hline
a15851 &5 &21      &15  & 51    \\ \hline \hline
a1412  &1  &17      &16  & 54     \\ \hline
a1434  &5  &20      &18  & 44     \\ \hline
a15500 &2  &18     &10  & 45    \\ \hline \hline
a16618 &5  & 23     &20  & 50    \\ \hline
a16628 &6 &26      &26  & 60    \\ \hline \hline
b389   &5  &18      &11  & 36    \\ \hline
b404   &8  &27    &16  & 52     \\ \hline

      \end{tabular}
    \end{center}
  \end{table}

We also observed striking differences in recovery of the initial
 optical transmittance at the room and low temperatures.
Almost no recovery is observed at -25~$^\circ$C
(see the first time period indicated in Fig.~\ref{fig:recovery}).
 Meanwhile, when the temperature is fixed at +20~$^\circ$C,
we can see a noticeable growth of the crystal transmittance
 (the second time period in Fig.~\ref{fig:recovery}).

The results of irradiation over 80 hours of the ten crystals
at 2 and 20 rad/h dose rate for the two temperatures,
 +20~$^\circ$C and -20~$^\circ$C, are collected
 in the Table~\ref{tab:summary2}.
  We have not seen any significant difference in radiation hardness
  of the ten crystals dependent on the doped element (see Tables 
\ref{tab:summary} and \ref{tab:summary2}) neither on the manufacturer. 
Though we might notice that crystal a15608 is the most radiation hard.

\section{Discussion}
It is most probable that the increase of light output at low
 temperatures is caused by decreased loss of excitation via
 non-radiative recombination.
Meanwhile, the lower radiation hardness and suppressed recovery
 after irradiation,
 which were observed 
 at low temperatures, are of considerable
 importance in application of PWO scintillation crystals as scintillators
 in cooled detectors. 
As it is generally accepted, the changes of light output of PWO  crystals
 subjected to irradiation are caused not by any variations  in
 emission efficiency but rather by radiation-induced absorption.
 Peculiarities of these features of radiation hardness  will be studied
 in more detail in further publications.
 There are currently two approaches explaining the optical properties
 of PWO crystal under irradiation~\cite{ann,bur1,bur2}. 

According to one approach~\cite{ann}, transmission damage occurs in
 crystal when  valence electrons are trapped in metastable states around
 the crystal defects.
 The irradiation of the crystals creates color centers which absorb
 light in particular spectral regions and cause reabsorption of the PWO
 emission.  When the rate of color centers production
 (proportional to the irradiation dose rate)
equals the natural temperature-dependent recovery rate, the crystal
 light output reaches a saturation level (quasi-plateau).
 Thus, the temperature dependent features of irradiation-induced absorption
 and recovery of the initial transmittance after termination
 of the irradiation can be interpreted by the shift of the balance
 between production and recovery of the color centers. 

According to another approach~\cite{bur1,bur2},  the results presented
 above are in agreement  with interpretation of radiation-induced
 changes in light output  by structural changes in inclusions
 of variable-valency tungstate  oxides $WO_{3-x}$~\cite{bur2}.
  Since tungsten has variable valency, irradiation  invokes changes
 in tungsten valency, rearrangement of oxygen  ions in
 the clusters $WO_{3-x}$, and, consequently,  in change of absorption
 by these clusters.  The irradiation-induced changes in the structure
 of $WO_{3-x}$  clusters proceed slower at decreased temperatures,
 and the thermal energy is insufficient to facilitate the rearrangement
 of the initial cluster structure, which is necessary for recovery of
 initial optical transmittance. 

\section{Summary}
To the best of our knowledge, this is the first study of radiation
hardness of PWO scintillation crystals at temperatures decreased below
 the room temperature. Using a selection of 10 similar commercial PWO
 crystals produced by different manufactures we demonstrate that
 the radiation-induced decrease of light output proceeds slower at
 lower temperatures and the initial output does not recover after
 termination of irradiation when the crystal temperature is maintained
 at -25~$^\circ$C. Nevertheless, the constant value of light output,
 which is reached after certain time under crystal irradiation
 at -25~$^\circ$C, is still considerable higher
 than that at the room temperature.
 Thus, cooling of PWO detectors is beneficial, though the long time
 evolution of the light output of the crystal under irradiation at
 decreased temperature (of the order of hundreds hours) should be
 taken into account in application of these detectors.

\section{Acknowledgments}
 This work has been partially supported by the INTAS grant N2 06-10000012-8845.
 We would like to thank the management of IHEP Radiation Research
Department for providing us a $^{137}Cs$ radioactive source for
our gamma irradiation studies.

\end{document}